\documentclass[a4paper,fleqn,donfloats]{cas-dc}

\usepackage[numbers,sort&compress]{natbib}
\usepackage{siunitx}
\usepackage{graphicx} 
\usepackage{subcaption}
\usepackage{float} 
\usepackage{placeins}
\usepackage{caption}
\begin{document}
\let\WriteBookmarks\relax
\def\floatpagepagefraction{1}
\def\textpagefraction{.001}


\shorttitle{Liquid Handling of the JUNO Experiment}

\shortauthors{J. Li et~al.}

\title [mode = title]{Liquid Handling of the JUNO Experiment}                      

\author[1]{Jiajun Li}[style=chinese]


\affiliation[1]{organization={Sun Yat-sen University},
    city={Guangzhou},
    postcode={510275}, 
    country={China}}

\author[2]{Yuekun Heng}
\cormark[1]

\affiliation[2]{organization={Institute of High Energy Physics},
    city={Beijing},
    postcode={100049}, 
    country={China}}

\author[1]{Jiajie Ling}
\cormark[2]

\author[2]{Zhi Wu}
\cormark[3]

\author[2]{Xiao Tang}
\author[2]{Cong Guo}
\author[2]{Jinchang Liu}
\author[2]{Xiaolan Luo}
\author[2]{Xiao Cai}
\author[1]{Chengfeng Yang}
\author[2]{Xiaoyan Ma}
\author[2]{Xiaohui Qian}
\author[1]{Tao Huang}
\author[1]{Bi Wu}
\author[1]{Pengfei Yang}
\author[1]{Shiqi Zhang}
\author[1]{Baobiao Yue}
\author[1]{Shuaijie Li}
\author[3]{Lei Yang}
\author[2]{Mei Ye}
\author[2]{Shenghui Liu}

\affiliation[3]{organization={Dongguan University of Technology},
    city={Dongguan},
    postcode={523000}, 
    country={China}}

\cortext[cor1]{Corresponding author. E-mail address: hengyk@ihep.ac.cn}
\cortext[cor2]{Corresponding author. E-mail address: lingjj5@mail.sysu.edu.cn}
\cortext[cor3]{Corresponding author. E-mail address: wuz@ihep.ac.cn}

\begin{abstract}
The Filling, Overflow, and Circulation (FOC) system is a critical subsystem of the Jiangmen Underground Neutrino Observatory (JUNO), responsible for the safe handling of the Liquid Scintillator (LS) and water throughout the detector's commissioning and operational lifetime. This paper details the design and operation of the FOC system, which accomplished the filling of the world's largest LS detector—taking 45 days for water (\SI{6.4e4}{\cubic\meter}) and 200 days for LS (\SI{2.3e4}{\cubic\meter}). Throughout water filling, the liquid level difference between the Central Detector and Water Pool was rigorously maintained within safety limits. During LS filling, level control achieved $\pm$2 cm precision with flow regulation within $\pm$0.5\% of setpoints. An automated control system based on Programmable Logic Controllers and the Experimental Physics and Industrial Control System framework ensured reliable operation. The system preserved LS radiopurity, maintaining \(^{222}\)Rn below 1 mBq/m\(^3\) during filling and achieving \(^{238}\)U/\(^{232}\)Th concentrations below \(10^{-16}\) g/g. The successful commissioning and operation of the FOC system have established it as an indispensable foundation for the stable long-term operation of the JUNO detector.
\end{abstract}

\begin{highlights}
\item Successful filling of the world's largest LS detector (20 kton) for JUNO.
\item Automated control achieved $\pm$2~cm level and $\pm$0.5\% flow precision.
\item Maintained $^{222}$Rn $<$ 1~mBq/m$^3$ during filling; $^{238}$U/$^{232}$Th $<10^{-16}$~g/g.
\item CD-WP level difference rigorously controlled within FEA safety limits.
\item System ensures long-term operation with overflow and purification functions.
\end{highlights}


\maketitle

\section{Introduction}

The Jiangmen Underground Neutrino Observatory (JUNO) is strategically positioned approximately 53 km from the Taishan and Yangjiang Nuclear Power Plants~\cite{JUNOphy_det}. The primary physics objective of JUNO is to determine the neutrino mass ordering (NMO) through measuring the oscillated energy spectrum of reactor neutrinos emitted by the reactors at at 53 km away and to achieve sub-percent precision on oscillation parameters such as $\theta_{12}$, $\Delta m^{2}_{21}$, and $|\Delta m^{2}_{32}|$. Additionally, JUNO will study solar neutrinos, atmospheric neutrinos, supernova neutrinos, cosmic diffuse neutrinos, proton decay, etc, thus advancing research in these areas~\cite{NeuPhysJUNO}.

To achieve these physics objectives, the JUNO Central Detector (CD) uses 20~kton of Liquid Scintillator (LS) as its target mass, requiring an energy resolution of 3\% at 1~MeV and ultra-low radioactive background levels. The LS is contained within a 35.4-meter-diameter acrylic sphere (120 mm thick), which is supported by a 41.1-meter-diameter stainless steel structure, connected via 590 supporting bars. The entire CD is submerged in the high purity water of the Water Cherenkov detector and covered by a plastic scintillator array on the top, which together serve as a muon veto system. All components in contact with the LS are fabricated from low-radioactivity, LS-compatible materials, subjected to rigorous cleaning procedures to minimze background, and engineered with high air-tightness to prevent radon intrusion. For comprehensive details on JUNO's design and its updated physics goals, please refer to the JUNO Conceptual Design Report and recent publications~\cite{CD_paper}.

LS have been the preferred detection medium for many pioneering neutrino experiments~\cite{Cowan, LSreview, BOREXINO, kamland, DayaBay, RENO, SNO+} due to a combination of advantageous properties, including scalability, homogeneity, ease of purification, and cost-effectiveness. Furthermore, they offer high light yield and detection efficiency, which provide excellent sensitivity. To satisfy the stringent demands of JUNO, given its massive detector size, the LS must achieve an attenuation length greater than 20 meters while maintaining exceptional radiopurity. The key radioactive contaminants and their maximum allowed concentrations are summarized in Table~\ref{table:radio-contaminants}; for instance, concentrations of $^{238}\text{U}$ and $^{232}\text{Th}$ must be below $10^{-15}$ g/g for reactor neutrino studies, and even lower for solar neutrino analyses. To maximize scintillation light generation and propagation, the optimal LS composition employs linear alkyl benzene (LAB) as the solvent, loaded with \(2.5 \, \text{g/L}\) of diphenyloxazole (PPO) as the fluor and \(3 \, \text{mg/L}\) of bis(2-methylstyryl)benzene (bis-MSB) as the wavelength shifter~\cite{LSrece}.

The JUNO LS production and purification system was meticulously designed and constructed to fulfill these specifications, capable of providing purified LS at a rate of \(7\ \text{m}^3/\text{h}\) for the CD LS filling. The system encompasses multiple integrated plants: a \SI{5000}{\cubic\meter} storage tank for raw LAB and transportation pipeline system for raw LAB, connecting the surface and underground laboratories; an Alumina Filtration Plant~\cite{Al} for optical purification; a Distillation Plant~\cite{DisandStr} for removing heavy metals such as uranium and thorium; and a Mixing Plant~\cite{mixing} for compounding PPO and bis-MSB with LAB into the master solution, which involves significant purification of PPO before dilution into the final LS. After mixing, the LS is transferred to the underground hall via an inclined shaft pipeline. Subsequent purification stages in the underground hall include a Water Extraction Plant~\cite{waterex} for removing water-soluble elements (e.g., K, Pb, Ra) and a Stripping Plant~\cite{DisandStr} for extracting gaseous elements (e.g., Rn, Ar, Kr). The system is also equipped with the Online Scintillator Internal Radioactivity Investigation System (OSIRIS)~\cite{OSIRIS} for real-time radioactivity assay, and supported by auxiliary High-Purity Water and Nitrogen Plants to supply essential utilities for the purification processes.

\begin{table*}[htbp]
    \centering
    \caption{List of the main radio-contaminants in JUNO LS~\cite{NeuPhysJUNO}}
    \label{table:radio-contaminants}
    \begin{tabular}{lcc}
        \hline
        Radioisotope & Contamination source & JUNO minimum requirements \\
        \hline
        \(^{222}\text{Rn}\) & Air, emanation from material  & \(< 5\ \text{mBq}/\text{m}^3\) \\
        \(^{238}\text{U}\) & Dust suspended in liquid  & \(< 10^{-15}\ \text{g/g}\) \\
        \(^{232}\text{Th}\) & Dust suspended in liquid  & \(< 10^{-15}\ \text{g/g}\) \\
        \(^{40}\text{K}\) & Dust suspended in liquid, PPO  & \(< 10^{-16}\ \text{g/g}\) \\
        \(^{39}\text{Ar}\) & Air & \(< 50\ \mu\text{Bq}/\text{m}^3\) \\
        \(^{85}\text{Kr}\) & Air & \(< 50\ \mu\text{Bq}/\text{m}^3\) \\
        \hline
    \end{tabular}
\end{table*}

A specialized Filling, Overflow, and Circulation (FOC) system has been designed and implemented to accomplish the precise filling and long-term stable operation of the ultra-pure LS. This system is responsible for several critical processes: the initial water filling, which involves the synchronous filling of the CD and the surrounding Water Pool (WP) to displace air—minimizing subsequent LS exposure—and to provide a final rinse of the CD's inner surface; the subsequent LS filling operation, which systematically replaces the water with LS to fill the CD—enabling the detector's primary data-taking function; the management of LS overflow to accommodate volume changes induced by temperature fluctuations during long-term operation; and finally, providing the capability for online circulation and re-purification of the LS. To meet the exceptional demands of the JUNO experiment, the FOC system incorporates stringent radiopurity controls, high-precision instrumentation, and a robust automated control system.

This paper provides a comprehensive overview of the FOC system's design, operational procedures, and performance. Section~\ref{sec:FOC system} details the system's design philosophy, radiopurity requirements, and architectural components. Section~\ref{sec:Control_Operation} describes the strategies and implementation for water filling, LS filling, and overflow management. Section~\ref{sec:conclusion} summarizes the system's results, achievements and outlines future perspectives.

\section{FOC System: Design and Requirements}
\label{sec:FOC system}

\subsection{Functional Overview and System Requirements}
\label{subsec:requirements}

The primary functions of the FOC system encompass four critical processes for the JUNO detector: 
(1)~the initial filling of the CD and WP with pure water; 
(2)~the subsequent exchanging water with LS in the CD; 
(3)~managing LS overflow to accommodate thermal expansion and contraction during long-term operation; 
and (4)~facilitating online LS circulation for re-purification if necessary. 
As the final stage in detector commissioning, the FOC system must execute these processes while rigorously preserving the structural integrity of the CD and the optical/radiopurity properties of the LS. This mandates a comprehensive set of system requirements:

\textbf{Ultra-low Background Contribution:} The FOC system's contribution to the overall radioactive background in the LS must be minimal. A paramount constraint is the concentration of \(^{210}\text{Pb}\), which must not exceed \(10^{-24}\) g/g in the LS. The FOC system is allocated no more than 10\% of this total budget. This requirement drives the need for ultra-high purity materials and exceptional leak tightness throughout the system.

\textbf{Liquid Level and Flow Control:} Based on the primary function, the system must provide highly precise and reliable control of liquid levels and flow rates. This necessitates high-accuracy sensors and robust control logic to maintain level differences and flow setpoints within strict safety margins during all phases of operation.

\textbf{Automation and Operational Safety:} Full automation of the filling processes is required to ensure reproducibility and minimize human error. This includes the implementation of a comprehensive multi-mode alarm system and safety interlocks (e.g., for level and pressure) to guarantee operational safety and equipment protection.

\textbf{Material Purity and Leak Tightness:} All components in contact with the LS or water, including stainless steel tanks, pipes, and valves, must meet stringent purity and sealing standards. The \(^{238}\text{U}\) concentration must be below 0.7 ppb, with an inner surface roughness \(\leq 0.4\ \mu\text{m}\) (typically 0.2--0.3 \(\mu\text{m}\)) achieved through electropolishing. Leak tightness is critical for preventing radon ingress; individual components must achieve a leak rate less than \(10^{-8}\) mbar·L·s\(^{-1}\), while assembled systems and critical connections (e.g., chimney flanges) must maintain a leak rate less than \(10^{-6}\) mbar·L·s\(^{-1}\) throughout the 20-year operational lifespan.

\subsection{System Architecture and Components}

The architecture of the FOC system comprises several key subsystems, engineered to handle the large liquid volumes while adhering to the requirements outlined in Section~\ref{subsec:requirements}. Its core infrastructure includes:

\textbf{Storage and Overflow Tanks:} One primary LS storage tank and two LS overflow tanks, each with a volume of 50 m$^{3}$, serve as buffers during filling and circulation. They are crucial for compensating volume fluctuations caused by temperature changes, given the LS's thermal volume coefficient of $8.8 \times 10^{-4} \, ^\circ\text{C}^{-1}$.

\textbf{Piping, Chimney, Pumping and Valving Network:} A comprehensive network of pipelines, the top and bottom chimneys, and valves facilitates the transfer of water and LS. All valves and fittings are required to have a helium leak rate $< 10^{-6}\ \text{mbar} \cdot \text{L} \cdot \text{s}^{-1}$ to prevent radon ingress from underground air. Dedicated pump sets are employed for transferring water and LS. This includes electromagnetic pumps for LS and self-priming pumps for water, all configured with standby units for redundancy. All flanges are designed with double O-ring seals and the space between two O-rings is filled with protection nitrogen. All the connection points are sealed with gastight nitrogen boxes or aluminum foils as an additional layer of protection.

\textbf{Control System:} An automated control system, based on PLC and the EPICS framework, orchestrates all operations. It integrates high-precision sensors, actuators, safety interlocks, and data management to ensure precise and stable control. A detailed description is provided in Section~\ref{subsec:Controlsystem}.

\textbf{Ultra-Pure Nitrogen System:} A dedicated radon-free nitrogen system, comprising storage tanks, mass flow meters, and bubble bottles, provides a protective blanket over the FOC tanks, calibration house, and top chimney. This prevents the LS from exposure to oxygen, moisture, and atmospheric radon.

\begin{figure*}[ht]
 \centering
 \makebox[\textwidth][c]{\includegraphics[width=0.9\textwidth]{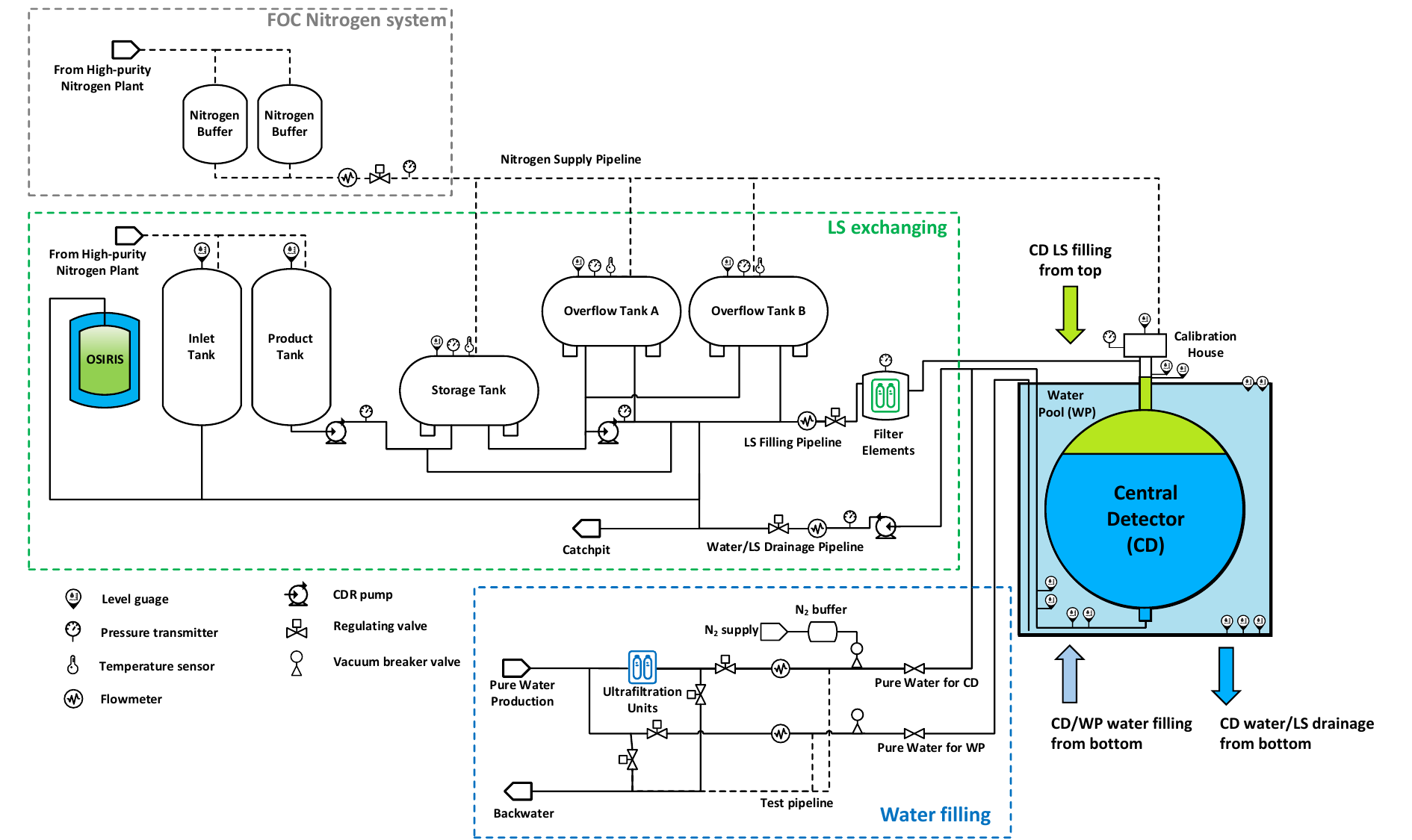}}%
 \caption{Schematic of the FOC system layout}
 \label{fig:foc_system}
\end{figure*}

The overall layout of the FOC system is depicted in Figure \ref{fig:foc_system}.

\subsection{Cleanliness and Quality Assurance}

Stringent cleanliness protocols are mandated for all FOC components that contact the LS or water. All internal surfaces undergo a meticulous multi-stage cleaning procedure to eliminate organic, inorganic, and particulate contaminants. The process sequentially includes: \textbf{degreasing} with high-purity detergents to remove oils, greases, and other organic residues; \textbf{acid pickling} with a dilute nitric acid solution to eliminate scale, oxides, and metallic contaminants from the activated surface; and \textbf{passivation} using concentrated nitric acid to form a stable, inert chromium oxide layer on stainless steel surfaces, thereby enhancing corrosion resistance and reducing future radon emanation. These operations are followed by a thorough rinsing with ultra-pure water (UPW) and drying under a high-purity nitrogen flush to prevent recontamination. After drying, components are properly sealed and stored under controlled conditions to maintain cleanliness. During installation, exposure to ambient air is minimized to prevent recontamination. The acceptance criteria for the final rinse water are rigorously defined, requiring compliance with the particle count specifications shown in Table~\ref{tab:cleanliness-specs} and achieving radiochemical contamination levels for \(^{238}\text{U}\) and \(^{232}\text{Th}\) below \(10^{-14}\) g/g. These measures are essential to prevent the introduction of particulate or radioactive contaminants that could jeopardize the detector's background levels.

\begin{table}[ht]
\centering
\caption{JUNO Requirements for the cleanliness of rinsed water~\cite{MIL-STD-1246C}}
\begin{tabular}{ccc}
\toprule
Particle Size & Surface density & Volume density \\
($\mu$m) & (counts/$0.1$ m$^2$) & (counts/L) \\
\midrule
5 & 179.0 & 1660 \\
15 & 27.0 & 250 \\
25 & 7.88 & 73 \\
50 & 1.08 & 10 \\
\bottomrule
\end{tabular}
\label{tab:cleanliness-specs}
\end{table}

\subsection{Automatic Control System}
\label{subsec:Controlsystem}

The FOC automatic control system~\cite{FOCcontrol} is a core technical component, ensuring the safe filling and 20-year stable operation of the 20-kton LS in the CD. Based on the ISA-88 standard~\cite{isa88}, it adopts a four-layer architecture (Sensor, Controller, Actuator, and Alarm \& Data Management Layers), integrating high-precision hardware and advanced control logic to meet JUNO's physics requirements.

\begin{figure*}[ht]
\centering
\includegraphics[width=0.65\textwidth]{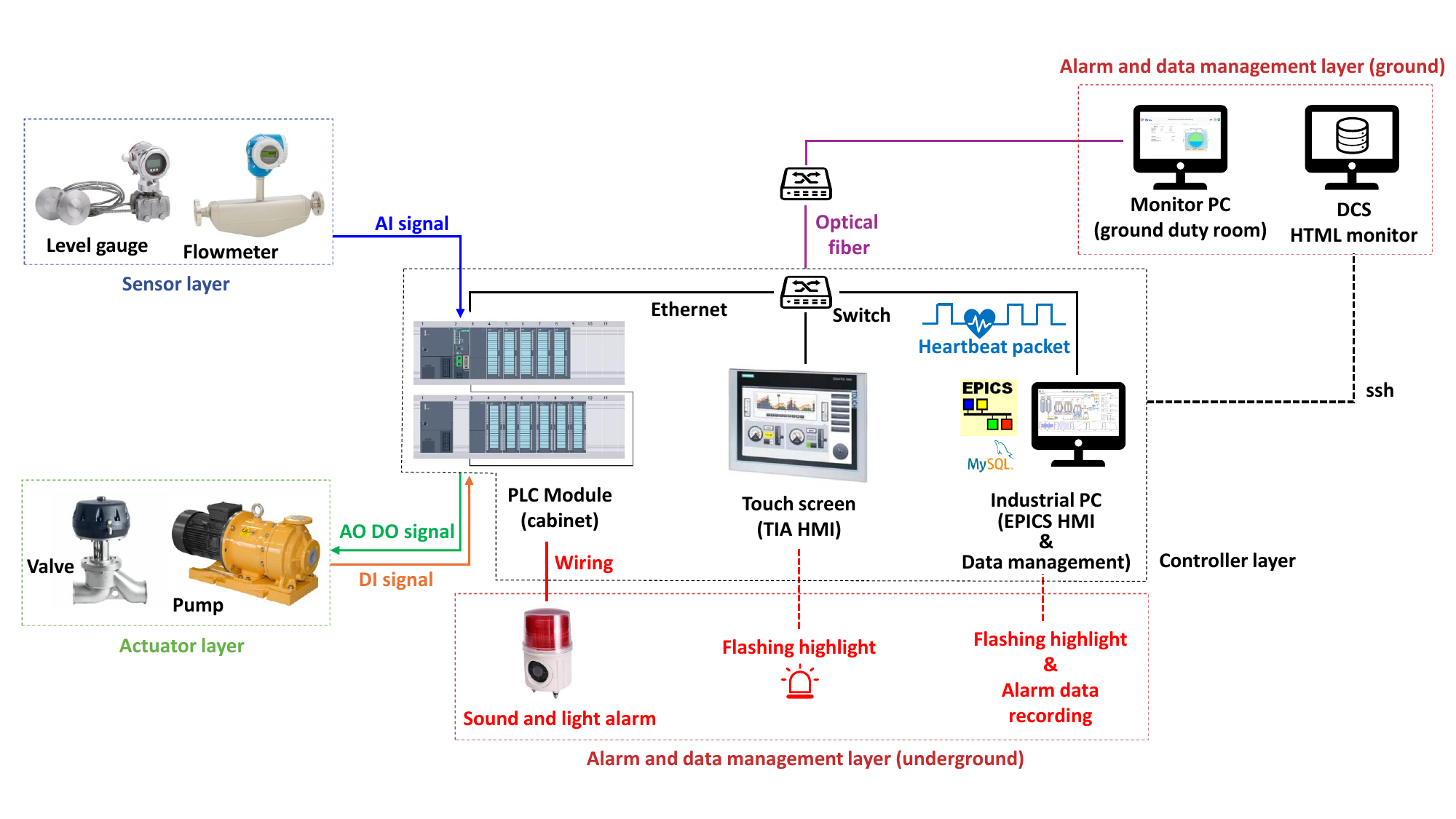}
\caption{FOC automatic control system}
\label{fig:Automatic_control}
\end{figure*}

\textbf{Sensor Layer}: All sensors (Endress+Hauser~\cite{endress_website}) feature an accuracy better than 0.2\% of full scale (FS). Redundancy is implemented for key parameters: the CD level is monitored by 2 differential pressure level gauges (0.20\% FS), 1 laser gauge (0.10\% FS), and 2 chimney differential pressure level sensors; the WP level uses 5 static pressure gauges; tank levels employ dual sensors (differential pressure + radar, 0.10--0.20\% FS). Shielded twisted-pair cables, LC filtering, isolators, and hybrid software filtering are utilized to suppress electromagnetic interference.

\textbf{Controller Layer}: A S7-300 PLC with microsecond-level response forms the core of this layer. Control software is developed on TIA Portal~\cite{TIA} for Human-Machine Interface (HMI) based visualization and integrated with EPICS~\cite{EPICS} for cross-system data exchange and real-time status monitoring.

\textbf{Actuator Layer}: This layer includes valves (SED\cite{SED}, leak rate $< 10^{-6}\ \text{mbar} \cdot \text{L} \cdot \text{s}^{-1}$, regulating valve accuracy $\pm 1\%$, on/off valve response $< 1\ \text{s}$) and pumps (8 units: CDR\cite{CDR} electromagnetic for LS, Yamei\cite{Yamei} self-priming for water, all with standby units). Real-time status monitoring and protection interlocks (e.g., level, pressure) are implemented.

\textbf{Alarm and Data Management Layer}: This layer is responsible for system-wide supervision, event logging, and ensuring operational safety. A multi-mode alarm system integrates on-site audiovisual alerts for immediate operator awareness, real-time highlighting on the HMI, remote notifications for off-site monitoring, and independent hardware emergency stop buttons for critical safety interventions. All operational data—including sensor readings, valve states, and alarm histories—are stored locally with high reliability and simultaneously synchronized to the Detector Control System (DCS) for distributed access and long-term archiving.

The integrated control system employs a combination of strategies, including Proportional-Integral-Derivative (PID) control for continuous parameters like flow rate, sequential control for complex operational procedures, and comprehensive safety interlocks. Commissioning tests confirmed the flow or liquid level control accuracy of 0.5\%, fully meeting JUNO's stringent specifications. The architecture of the automatic control system is illustrated in Figure~\ref{fig:Automatic_control}.

\section{Liquid Management Design and Operation}
\label{sec:Control_Operation}

This section provides a comprehensive overview of the control strategies, operational execution, and performance outcomes for the key liquid handling processes managed by the FOC system: water filling, LS filling, and LS overflow management. For each process, we first describe the underlying control logic and design considerations, followed by a detailed account of the operational implementation, challenges encountered, and the final achieved results. This integrated presentation highlights the seamless transition from theoretical control principles to practical execution, demonstrating the system's robustness and adaptability in meeting JUNO's stringent requirements.

\subsection{Water Filling}

\subsubsection{Control Strategy and Logic}
To minimize LS exposure to air, the air inside the CD must first be replaced with water or ultrapure nitrogen, followed by replacing the water or ultrapure nitrogen inside the CD with LS. To reduce engineering risks, the filling scheme of replacing water with LS is adopted.

The pure water production system, with a capacity of \SI{100}{\cubic\meter/\hour}, was developed to meet stringent requirements for detector filling and operation. The final product water must satisfy radiopurity criteria including $^{222}\text{Rn} < \SI{10}{\milli\becquerel\per\cubic\meter}$, $^{226}\text{Ra} < \SI{50}{\micro\becquerel\per\cubic\meter}$, and $^{235}\text{U}$/$^{232}\text{Th}$ concentrations below $\SI{e-15}{\gram\per\gram}$, along with physicochemical specifications of \SI{20}{\celsius} temperature, dissolved oxygen $< \SI{10}{ppb}$, and resistivity $> \SI{18}{\mega\ohm\cdot\centi\meter}$. The purification employs a two-stage process. Aboveground pretreatment includes bag filtration, multi-media filtration, activated carbon adsorption, radium-removal softening resin columns, and primary Reverse Osmosis (RO). The water is then transported via a \SI{1300}{\meter} stainless-steel pipeline to the underground facility for further polishing through secondary RO, Electrodeionization (EDI), \SI{0.1}{\micro\meter} cartridge filtration, UV sterilization, and advanced radon removal using a five-stage degassing membrane system augmented by microbubble technology. This configuration achieves a radon removal efficiency exceeding \SI{99.9}{\percent}, reducing concentrations to $\SI{0.61 \pm 0.50}{\milli\becquerel\per\cubic\meter}$ in recirculation mode. For CD filling, an additional Ultrafiltration (UF) step is implemented, with approximately \SI{5}{\cubic\meter\per\hour} water loss during this process. The system operates in both filling and recirculation modes to support initial commissioning and long-term detector operation~\cite{PWprod}.

Prior to water filling, the Acrylic Spheroidal Vessel (ASV) undergoes comprehensive cleaning to meet ultra-high purity standards. This process includes reducing airborne dust via water mist spraying, mitigating radon through natural decay, and stripping protective films/cleaning the acrylic inner surface with high-pressure rotary nozzles. Those films were applied during the ASV installation to prevent airborne radon from penetrating the acrylic material, while also providing an additional protective layer for the acrylic. Concurrently, the commissioning pipeline tests and debugs the water filling automatic control system to ensure reliability.

\vspace{8pt} 
\noindent
\begin{minipage}{\columnwidth}
    \centering
    \includegraphics[width=0.9\columnwidth]{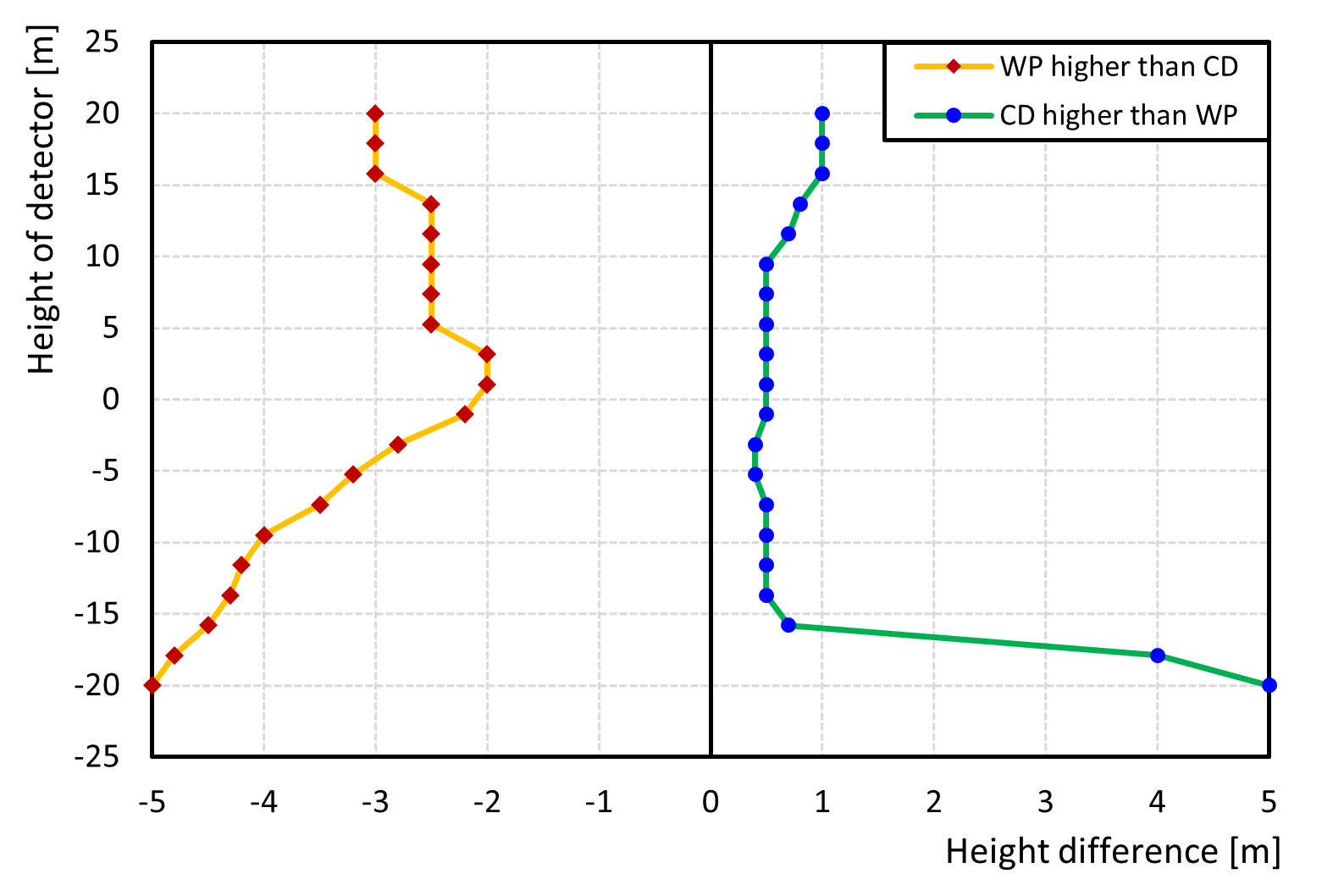}
    \captionof{figure}{Limit of the height difference ($\Delta H$) between the CD and WP during water filling~\cite{FEA}}
    \label{fig:Limit_diff}
\end{minipage}
\vspace{8pt} 

The main challenge in synchronizing the CD and WP water filling was managing their differing cross-sectional areas, which vary with height. The paramount objective was to maintain the liquid level difference ($\Delta H$, defined as $H_{\text{CD}} - H_{\text{WP}}$) strictly within the safe limits predetermined by finite element analysis (FEA)~\cite{FEA}. As visualized in Figure \ref{fig:Limit_diff}, a larger allowable range on the left (negative $\Delta H$), which indicates that the spherical acrylic vessel can withstand a higher external pressure from the WP than internal pressure. Consequently, a slightly higher WP level was permitted during the water filling process. The implemented control strategy was multifaceted and relied on several key pillars:

\begin{itemize}
    \item \textbf{WP Level as Primary Reference}: The real-time WP level, being highly accurate and stable, served as the primary control parameter. Accurate assessment of the CD level, susceptible to dynamic flow interference, required periodic intentional pauses in filling.
    
    \item \textbf{Guidance Flow Rate Calculation:} The required flow rates into the CD and WP branches were continuously calculated based on the precisely known geometries of the CD sphere, the WP cylinder, and all internal structures (such as PMTs, electronics boxes, and the support grid), combined with the total available water flow rate (\(\sim 90\ \text{m}^3/\text{h}\)).
    
    \item \textbf{PID Control for Flow Distribution:} Two sets of pneumatically actuated regulating valves, one for the CD branch and another for the WP branch, were controlled by PID algorithms. These valves automatically adjusted their opening to distribute the inlet flow, striving to achieve synchronized rising of the liquid levels in both vessels.
    
    \item \textbf{Multi-Tiered Compensation Logic:} If the measured level difference ($\Delta H$) indicated that one vessel was lagging, the guidance flow rate for that branch was increased in predefined incremental steps (1\%, 3\%, or 5\%) to expedite its filling and correct the imbalance.
    
    \item \textbf{Stringent Safety Interlock:} As a paramount protective measure, if $\Delta H$ exceeded a threshold set at one-quarter of the absolute design limit (providing a significant safety margin), a strict "on-off" control mode was automatically activated. In this mode, filling to the vessel with the higher level was completely halted, and flow was directed exclusively to the other vessel until the level difference was reduced to a safe value.
    
    \item \textbf{Special Handling at Critical Structural Regions:} Recognizing the non-linear volume changes and heightened mechanical sensitivity near the detector's South Pole, North Pole, equatorial region, and chimney interfaces, the control logic incorporated tailored measures for these zones. This included significantly reduced flow rates or the exclusive use of the cautious "on-off" filling mode to absolutely ensure mechanical safety and prevent excessive stress on the acrylic vessel.

    \item \textbf{Mitigation of Liquid Column Separation:} The significant $\sim$44 meter static head loss in the vertical filling pipelines created a risk of column separation. This phenomenon can be visualized as the water column undergoing a free fall within the pipe, which can tear the continuous liquid apart, creating discrete segments and pulling a vacuum between them. Physically, this occurs when the local pressure drops below the vapor pressure of the liquid, leading to the formation of vapor cavities. The subsequent collapse of these cavities can generate destructive water hammer. To mitigate this, vacuum break valves were installed on both the CD and WP filling lines. For the CD line, where the introduction of air was strictly prohibited to maintain radiopurity, a \SI{1500}{\liter} nitrogen buffer tank was implemented instead. This tank supplied high-purity nitrogen at a controlled rate of approximately \SI{15}{\cubic\meter\per\hour} to compensate for the pressure drop and prevent a vacuum, thereby ensuring safe and stable flow.
\end{itemize}

\vspace{8pt} 
\noindent
\begin{minipage}{\columnwidth}
    \centering
    \includegraphics[width=0.9\columnwidth]{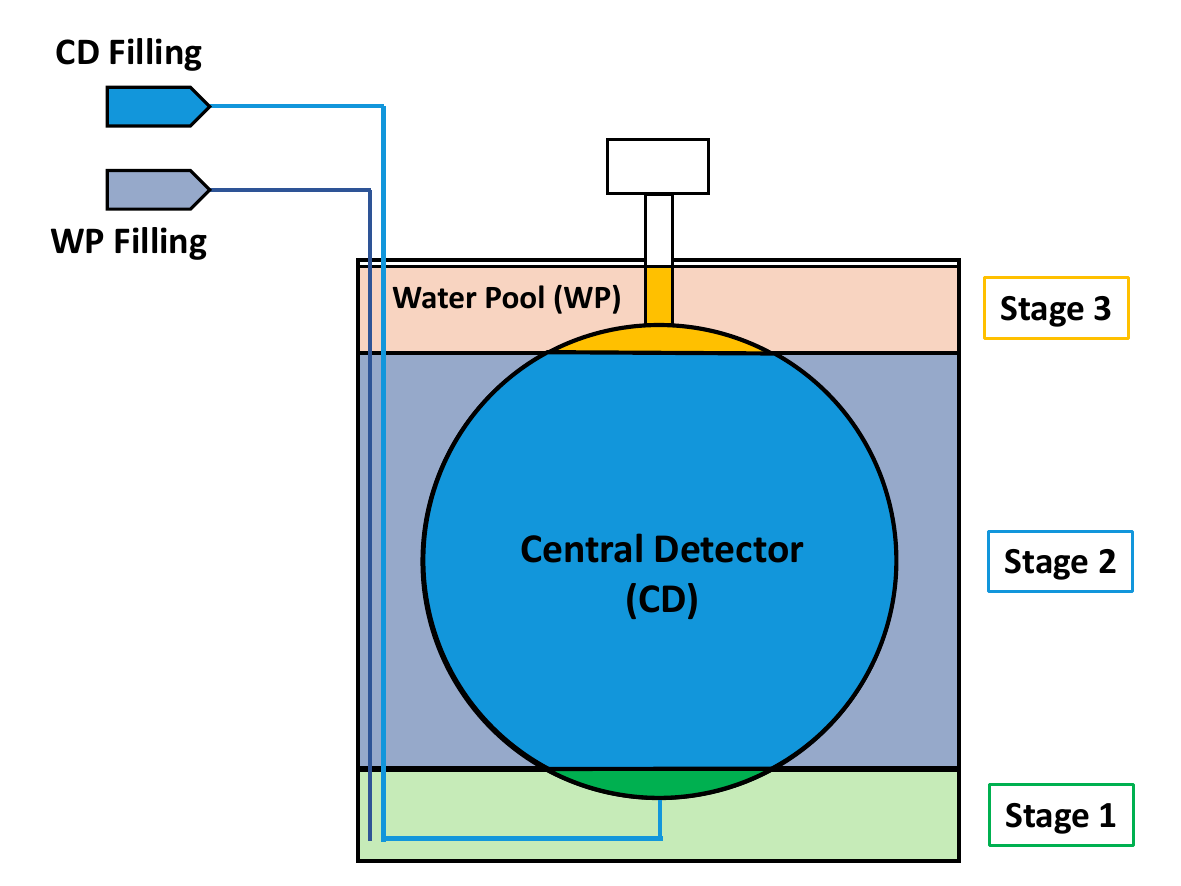}
    \captionof{figure}{Three filling stages of water filling}
    \label{fig:PW_filling_stages}
\end{minipage}
\vspace{8pt} 

\subsubsection{Operational Implementation and Results}

The water filling operation was meticulously executed in one pre-stage and three filling stages (Figure \ref{fig:PW_filling_stages}), each tailored to the specific geometric and mechanical challenges presented by different regions of the detector:

\textbf{Pre-stage - ASV Cleaning}: This initial stage was successfully completed, with all measured parameters exceeding the stringent predefined specifications. A pure water mist generator was employed to suppress airborne dust, significantly improving the internal air cleanliness from Class 100,000 to Class 100. Throughout the cleaning process, the distribution and flow rate of the rinse water were precisely managed by the FOC system. Particulate contamination levels and absorption spectra of the rinse water fully complied with the technical standards. Crucially, the cleaning procedure resulted in the near-complete removal of protective films, and the rinse water achieved exceptional ultralow radiochemical contamination levels.

\textbf{Stage 1 - South Pole Region}: This stage involved filling the WP volume below and at the South Pole while intermittently synchronizing the filling of the CD's South Polar region itself, the CD filling pipelines, and the bottom chimney. Given the extremely small volumes of the CD components in this area ($\sim200\ \text{m}^3$ for the South Polar region) compared to the WP, achieving synchronized level rise was exceptionally challenging. Consequently, a conservative "on-off" mode was employed. The WP was filled continuously at a high rate of $\sim75\ \text{m}^3/\text{h}$. After every 20-30 cm rise in the WP level, the filling process was briefly paused to allow for minimal-flow (1-2 $\text{m}^3/\text{h}$) topping-up of the CD-related components until their levels matched the new WP height. This method ensured successful and safe filling of this critical region.
    
\textbf{Stage 2 - Interpolar Region}: Covering the vast volume between the poles, this stage benefited from slower and more predictable changes in the cross-sectional areas of both the CD and WP. The CD filling pipeline could now sustain a stable controlled flow of $\sim8\ \text{m}^3/\text{h}$. The primary control logic involving PID-controlled valve adjustment and guidance flow rate calculation was deployed here. A total filling rate of $\sim80\ \text{m}^3/\text{h}$ was maintained. To accommodate necessary daily maintenance of the pure water production system (e.g., filter bag replacements) and to enable accurate verification of the level difference through periodic pauses, two scheduled daily reductions or interruptions in the total filling flow were incorporated, ensuring both operational efficiency and detector safety.

\vspace{8pt}
\noindent
\begin{minipage}{\columnwidth}
    \centering
    
    \includegraphics[width=0.9\columnwidth]{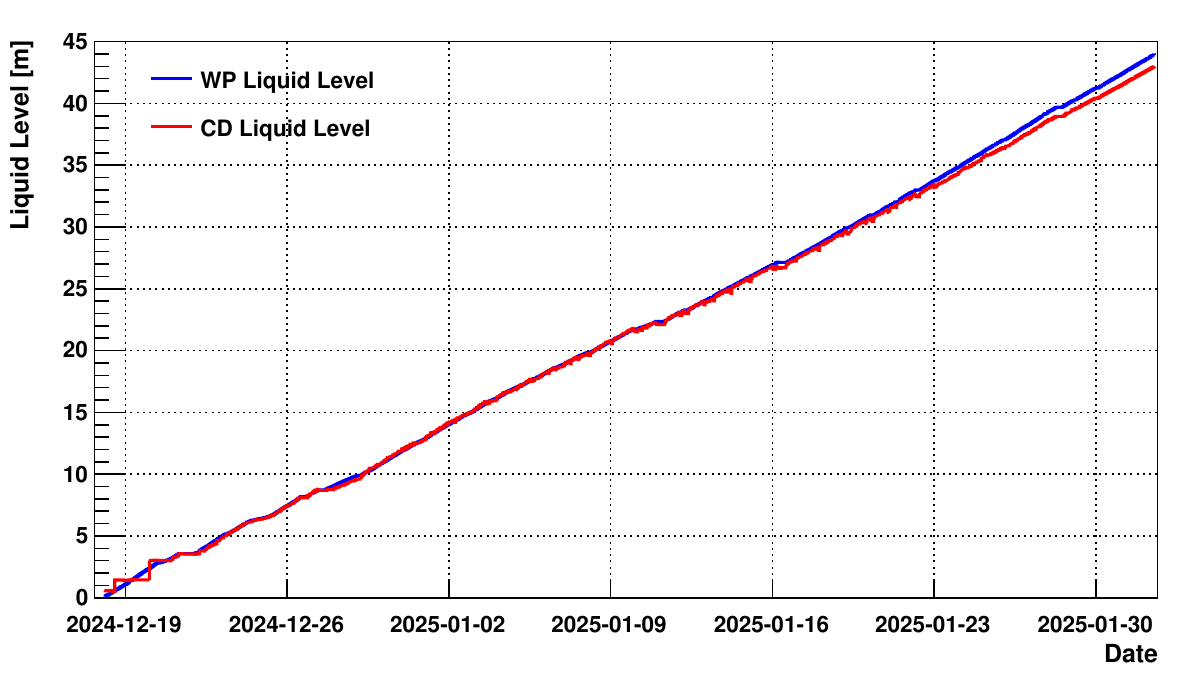} \\
    \vspace{3pt}
    \small (a) Water level changes
    \label{fig:PW_level_changes}
    
    \vspace{8pt}  
    
    \includegraphics[width=0.9\columnwidth]{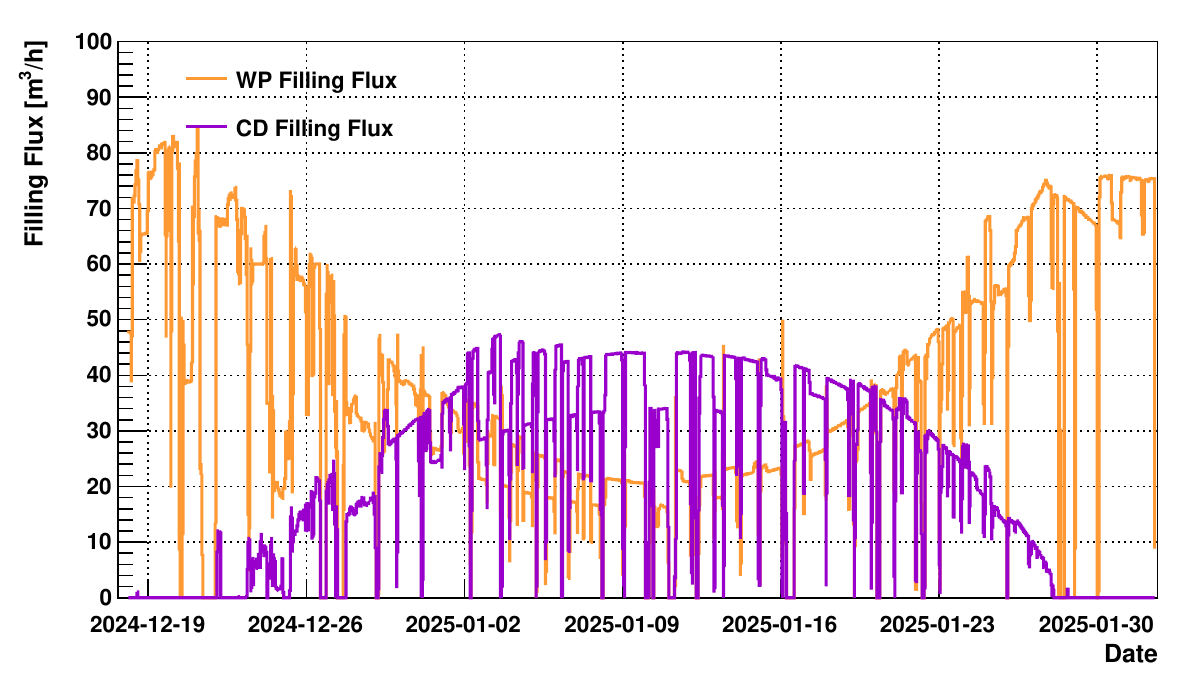} \\
    \vspace{3pt}
    \small (b) Filling flow changes
    \label{fig:PW_flow_changes}

    \vspace{5pt}
    \captionof{figure}{Level and flow changes during the water filling process}
    \label{fig:PW_filling_changes}
\end{minipage}
\vspace{8pt}
    
\textbf{Stage 3 - North Pole \& Top Chimney}: Similar to Stage 2, this final stage handled the North Polar region of the CD (volume $\sim200\ \text{m}^3$) and the top chimney separately from the remaining WP volume. The "on-off" mode was reinstated for the North Pole region. A significant and advantageous phenomenon was observed for the top chimney: with over $23{,}000\ \text{m}^3$ of water already inside the CD, when CD filling was paused and WP filling continued, the increasing hydraulic pressure from the rising WP level caused a slight elastic contraction of the CD sphere. This contraction, in turn, caused the water level in the top chimney to rise synchronously with the WP level, even without active CD filling. During actual operations, only minimal supplemental filling was required for final level adjustment.

The entire process of filling approximately \SI{6.4e4}{tons} of ultra-pure water was completed successfully within 45 days, commencing on December 18th, 2024, and concluding on February 1st, 2025. Throughout the operation, the flow rate was precisely controlled, with the actual flow maintained within $\pm 1\%$ of the theoretical target flow. The evolution of the water levels and filling flow rates throughout this period is depicted in Figure \ref{fig:PW_filling_changes}. Visual documentation of the filling process is provided in Figure \ref{fig:PW_filling_process}. At the very end of the detector water filling, on the far right side of the Figure \ref{fig:PW_filling_changes}, a minor calibration issue with the WP level gauge towards led to a deviation in the final WP level control. However, critically, the liquid level difference ($\Delta H$) throughout the entire filling period consistently remained within the safe design requirements stipulated by the mechanical analysis (Figure \ref{fig:Limit_diff}).

\vspace{8pt}
\noindent
\begin{minipage}{\columnwidth}
    \centering
    
    \includegraphics[width=0.8\columnwidth]{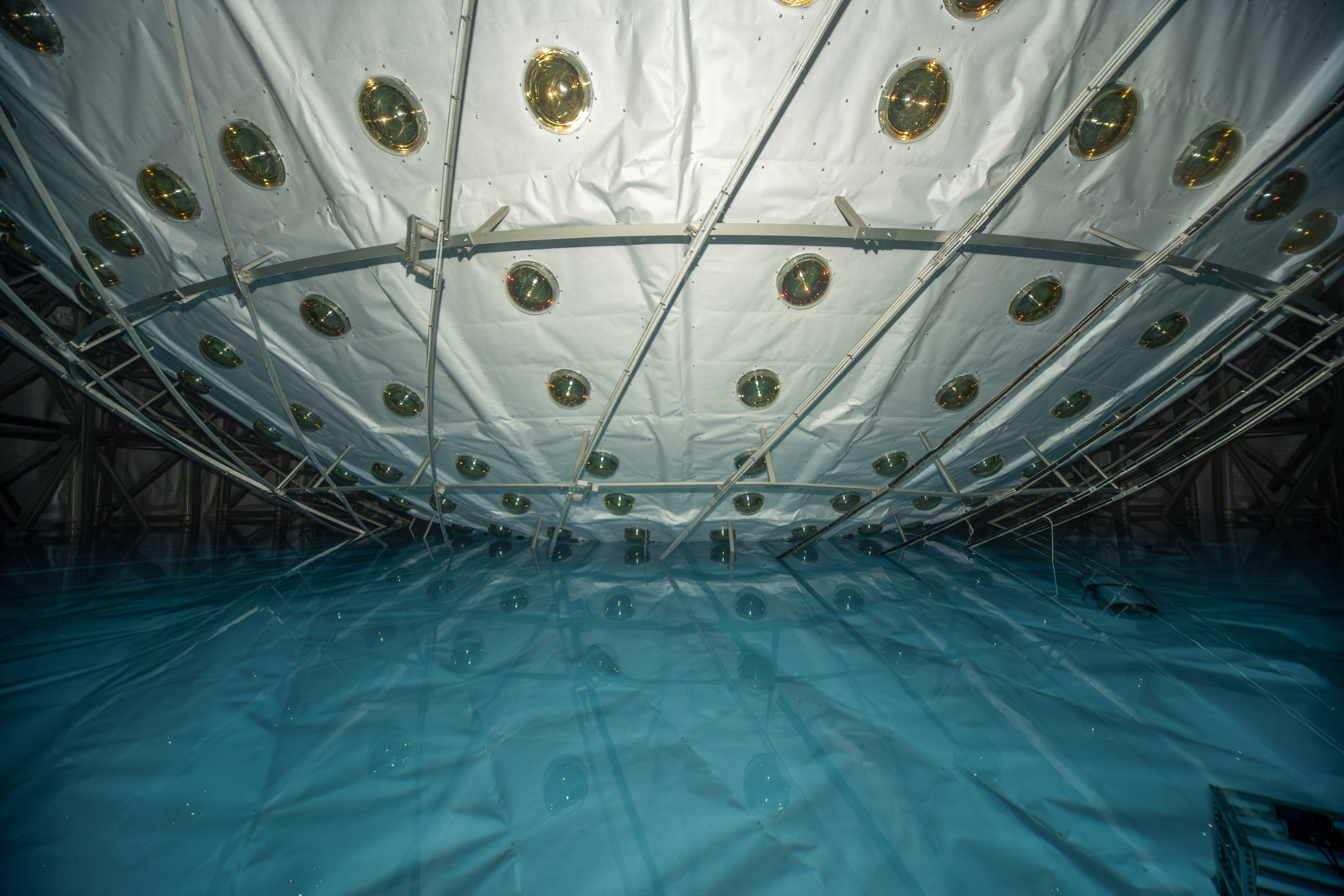} \\
    \small (a) Start of water filling (detector bottom)
    \label{fig:PW_filling_start}
    
    \vspace{8pt}
    
    \includegraphics[width=0.8\columnwidth]{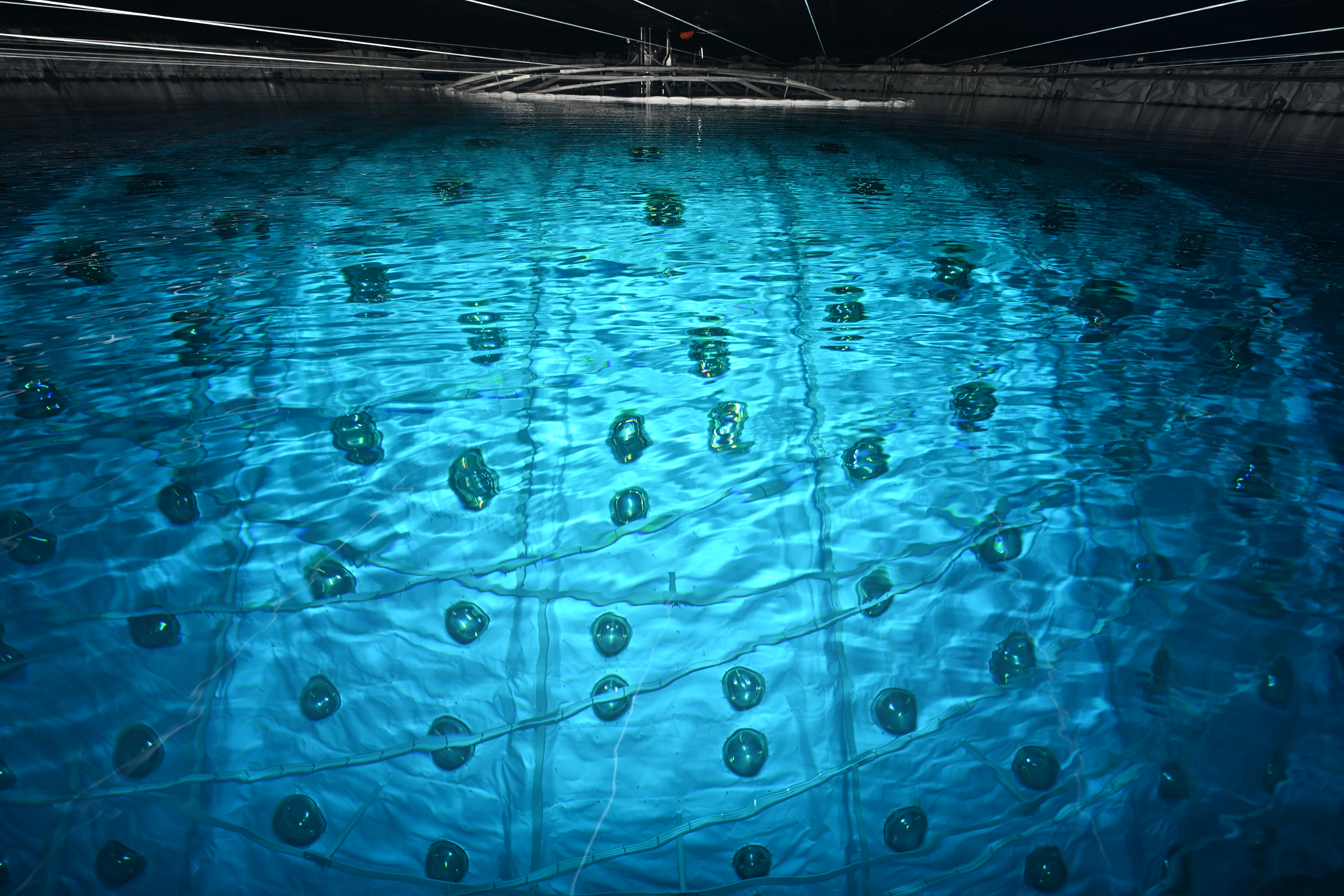} \\
    \small (b) Completion of water filling (detector top)
    \label{fig:PW_filling_end}
    
    \vspace{5pt}
    \captionof{figure}{Illustration of the water filling process}
    \label{fig:PW_filling_process}
\end{minipage}
\vspace{8pt}

\subsection{LS Filling}

\subsubsection{Control Strategy and Logic}
\label{sssec:ls-control-logic}

Following the successful completion of water filling, the CD interior was purged with ultrapure nitrogen that had a $^{222}\text{Rn}$ contamination level below $10\ \mu\text{Bq}/\text{m}^3$, the critical LS filling operation commenced. This process involved the continuous injection of purified LS into the CD from the top chimney while simultaneously draining the water from the bottom pipeline. The LS was supplied at a nominal rate of $7\ \text{m}^3/\text{h}$ by the dedicated JUNO LS production and purification system \cite{LSrece}, which encompasses nine subsystems including storage, alumina filtration, distillation, mixing, water extraction, gas stripping, and online radioactivity monitoring (OSIRIS), along with ultrapure water and high-purity nitrogen supply systems for comprehensive purification and quality assurance.

The primary challenge, central to the control strategy, arose from the significant density difference between LS ($\rho \approx 0.856\ \text{g/cm}^3$ at \SI{21}{\celsius}) and water. As the lighter LS progressively exchanged the pure water in the CD, the increasing buoyant force exerted on the submerged acrylic vessel raised the risk of mechanical failure. To mitigate this risk and manage the pressure differential across the acrylic vessel, the liquid level inside the CD was deliberately and incrementally raised during the filling process. This level adjustment aimed to shift the pressure equilibrium point close to the vessel's equator, with the goal of minimizing the resultant pressure differential across the acrylic vessel. The primary objective of this strategy was to reduce the maximum stress on the spherical structure throughout the exchange. Consequently, the control strategy was designed to maintain the pressure equilibrium point near the ASV's equator by gradually increasing the CD liquid level during the filling process. Ultimately, the design required the final LS level to be maintained several meters above the external WP water level to ensure a safe and stable pressure differential throughout the detector's operational lifetime. 

The control logic for the LS filling was designed to manage this dynamic process automatically and safely:
\begin{itemize}
    \item \textbf{Real-time Level Guidance and LS/Water Interface Tracking}: The control system relied on multiple redundant sensor inputs. Two sets of differential pressure level gauges installed on the CD chimney provided the primary measurement of the LS surface level. The position of the LS-water interface inside the CD was tracked in real-time based on the total volume of LS injected and water drained, and was continuously validated against measurements from four additional CD level gauges and independent float level gauge readings.
    
    \item \textbf{Flow Control and Balancing}: The primary control objective was to maintain the LS surface level within \(\pm 20\ \text{cm}\) of its expected value at any given stage. The water drainage flow rate was typically fixed at the nominal production rate of \(7\ \text{m}^3/\text{h}\). The LS injection flow rate was then finely adjusted around this baseline value using regulating valves and auxiliary pump frequency control, based on the real-time discrepancy between the actual and target LS levels.
    
    \item \textbf{Periodic Recalibration and Phase Management}: Recognizing potential deviations from ideal conditions (e.g., slight mismatches in inflow/outflow, density variations), the LS filling process was programmed for periodic pauses. During these pauses, the current state of the LS filling (the current 'phase') was thoroughly reassessed based on the most accurate level measurements. The guidance flow rates for the subsequent segment were then adjusted accordingly, often involving compensatory increases or decreases in the LS injection rate to bring the levels back in line with the pressure differential limits defined for that specific stage of the operation.
    
    \item \textbf{Stringent Safety Interlock}: The utmost priority was the structural integrity of the ASV. The control system continuously monitored the calculated pressure differential across the acrylic vessel. If this differential approached a predefined maximum allowable threshold (derived from FEA), a paramount safety interlock was triggered. This interlock activated an "on-off" LS filling mode, which immediately halted either LS injection or water drainage (whichever action would most effectively reduce the pressure differential) until the condition returned to a safe range.
    
    \item \textbf{Coordination with LS Production}: The FOC system's extraction of LS from the product tank needed to be synchronized with the production plant's output rate of $7\ \text{m}^3/\text{h}$. In the absence of a dedicated flow meter on this transfer line, synchronization was achieved by periodically fine-tuning the transfer pump's frequency based on the measured liquid levels in both the LS product tank and the FOC storage tank, ensuring a dynamic balance between production and consumption for the LS filling.
    
    \item \textbf{Backup of Water Drainage Capability:} A backup design consideration was the evolving hydraulic head for water drainage. As LS replaced the denser water and the LS/water interface descended, the static pressure difference between the CD and the WP gradually increased. This meant the self-priming drainage pumps effectively had to lift water from an increasingly greater depth, equivalent to a suction head rising from an initial 0.5 m to a design maximum of approximately 4 m. To safeguard against potential pump performance degradation or instability under this varying load, a vacuum-assisted backup system was engineered. Although this system was never activated during actual operation, as the primary drainage system maintained stable performance throughout the entire LS filling process, its design and integration provided a critical safety margin to ensure the continuous outflow rate of $7\ \text{m}^3/\text{h}$ under all anticipated conditions.
\end{itemize}

\subsubsection{Operational Implementation and Results}
\label{sssec:ls-op}

The LS filling operation was executed in a carefully sequenced manner, prioritizing both system verification and radiopurity assessment:

\textbf{Initial Batch and Purity Check (Feb.8-10, 2025)}: The operation commenced with the successful injection of an initial batch of $100\ \text{m}^3$ of LS. This batch was then held within the CD for several days to acquire data. This waiting period allowed for the decay of any $^{222}\text{Rn}$ that was not in secular equilibrium with its parent nuclide and provided the first crucial opportunity to assess whether the intrinsic radiopurity of the newly produced scintillator met the experiment's stringent requirements using the OSIRIS system.

\textbf{Commencement of Continuous Filling (Feb.25, 2025)}: Following the positive initial assessment, the purification plants were configured for round-the-clock operation, and the continuous phase of LS filling began. A deliberately cautious approach was adopted initially: the filling rate was set to a low range of $1$--$2\ \text{m}^3/\text{h}$ for the first few hours. This allowed for a final verification of the stability and safety of the entire integrated supply chain—from production to injection—and the CD's response to the ongoing LS filling.

\textbf{Running-in Phase (Feb.25 - Mar.10, 2025)}: After the initial verification, the flow rate was promptly ramped up to its nominal value of $7\ \text{m}^3/\text{h}$. This period, lasting approximately two weeks, is identified as the running-in phase for the filling system supporting the LS filling. During this phase, continuous 24/7 operation could not yet be fully sustained due to teething issues and planned initial adjustments, resulting in an operational availability of approximately 50\%.

\textbf{Stable Operation Phase (Mar.11 - Aug.22, 2025)}: From March 11th onward, the system entered a period of highly stable operation for the LS filling. Near-continuous 24/7 filling was achieved with significantly improved reliability, boasting an operational availability of 90\%. Throughout this entire period, the control system demonstrated exceptional performance: the liquid level inside the CD was meticulously maintained within a variation of $\pm 2$ cm of the predetermined target level trajectory, and the LS filling flow rate was precisely controlled within an accuracy of $\pm 0.5\%$ of the setpoint.

Throughout the LS filling process, comprehensive radiopurity monitoring was implemented to ensure ultra-low background levels, consistent with the requirements for JUNO’s physics goals. Real-time \(^{222}\text{Rn}\) activity monitoring was achieved by tagging \(^{214}\text{Bi-}^{214}\text{Po}\) cascade decays in the CD data stream, complemented by periodic batch testing of 3-5 tons LS samples using the OSIRIS system. The average \(^{222}\text{Rn}\) contamination in the filled LS was measured to be less than \(1 ~\text{mBq/m}^3\), which is well below the design requirement of \(5 ~\text{mBq/m}^3\)~\cite{performancejuno}. This excellent performance was attributed to the stringent leakage control measures, continuous nitrogen purging of the filling system and CD chimney, and the high-efficiency radon removal in the LS purification chain. 

Furthermore, independent radiopurity monitoring was conducted using Inductively Coupled Plasma Mass Spectrometry (ICP-MS), a highly sensitive technique capable of detecting trace levels of \(^{238}\text{U}\) and \(^{232}\text{Th}\)~\cite{ICP-MS}. In this method, uranium and thorium are first extracted from the liquid scintillator via acid digestion, concentrated, and then introduced into the ICP-MS system for quantitative analysis. With stringent cleanliness control throughout sample preparation---including the use of ultra-pure reagents and meticulously cleaned labware---the method achieves recovery efficiencies close to 100\% and a detection limit at the sub-ppq level. The ICP-MS results confirmed that the concentrations of \(^{238}\text{U}\) and \(^{232}\text{Th}\) in the LS were maintained below the \(10^{-16} \, \text{g/g}\) level throughout the filling process. This provides direct evidence that the FOC system did not introduce contamination from these critical long-lived radioisotopes.


Following the completion of LS filling, an integrated radiopurity analysis of the LS within the CD fiducial volume was performed. The results confirmed that the concentrations of \(^{238}\text{U}\) and \(^{232}\text{Th}\) were successfully maintained below the \(10^{-16}\) g/g level, along with an initial \(^{210}\)Po rate of approximately \(5 \times 10^{4}\) cpd/kton~\cite{performancejuno}~\cite{junoosci}, fully meeting the stringent requirements for both neutrino mass ordering and solar neutrino studies. This demonstrates that the FOC system effectively prevented the introduction of external contaminants throughout the filling process, thereby preserving the ultra-high radiopurity of the LS.

The monumental task of LS filling \SI{2.3e4}{\cubic\meter} of water with LS was successfully concluded on August 22nd, 2025, after a total duration of 200 days. Following the main LS filling, an additional \(80\ \text{m}^3\) of LS was deliberately drained from the bottom of the detector. This volume, which constituted the initial batch filled into the CD, was purged due to its higher contamination risk from prolonged contact with residual water and particulates introduced during the initial pipeline flushing. This operation functioned as a one-time, unidirectional flushing cycle aimed at removing the potentially compromised scintillator from the detector's lowest point, thereby ensuring the final optimal purity within the CD. It is noteworthy that this process can be regarded as a specific form of circulation, albeit without routing the drained LS back through the external purification system. The control logic and operational procedures for this operation were fundamentally similar to those implemented for the online LS circulation described in Section \ref{subsec:ls-overflow}.

The downward migration of the water/LS interface and the corresponding increase in the total LS volume inside the CD throughout the LS filling process are clearly depicted in Figure \ref{fig:LSfilling_changes}. The flow rate stability during the stable phase is illustrated in Figure \ref{fig:ex_flux}.

\vspace{8pt} 
\noindent
\begin{minipage}{\columnwidth}
    \centering
    \includegraphics[width=1.0\columnwidth]{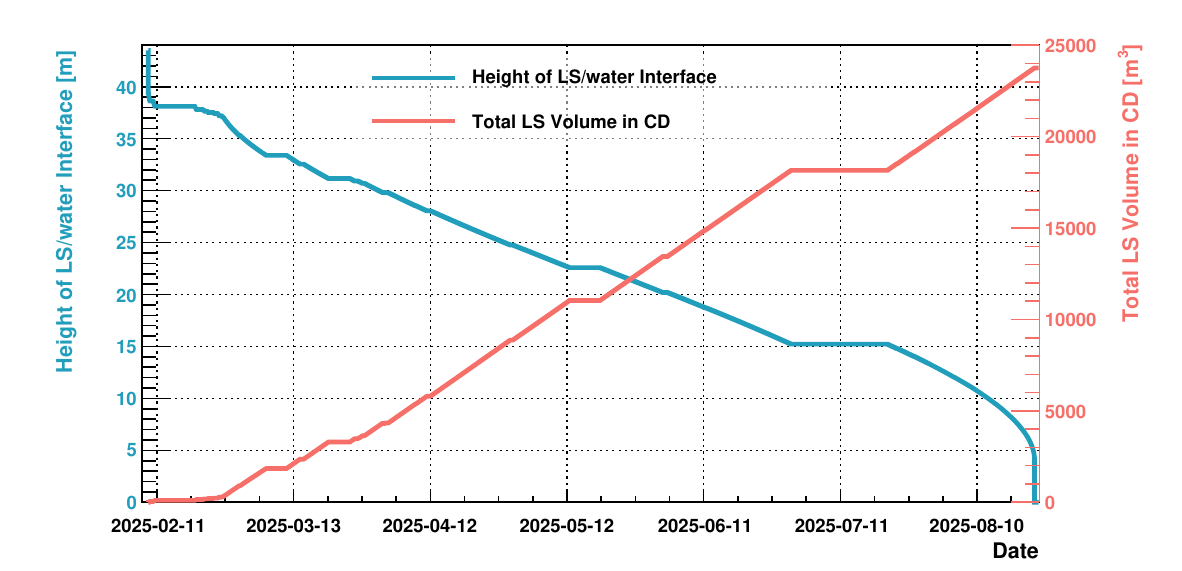}
    \captionof{figure}{LS/Water interface and LS volume changes during the LS filling process}
    \label{fig:LSfilling_changes}
\end{minipage}
\vspace{8pt} 

\vspace{8pt} 
\noindent
\begin{minipage}{\columnwidth}
    \centering
    \includegraphics[width=1.0\columnwidth]{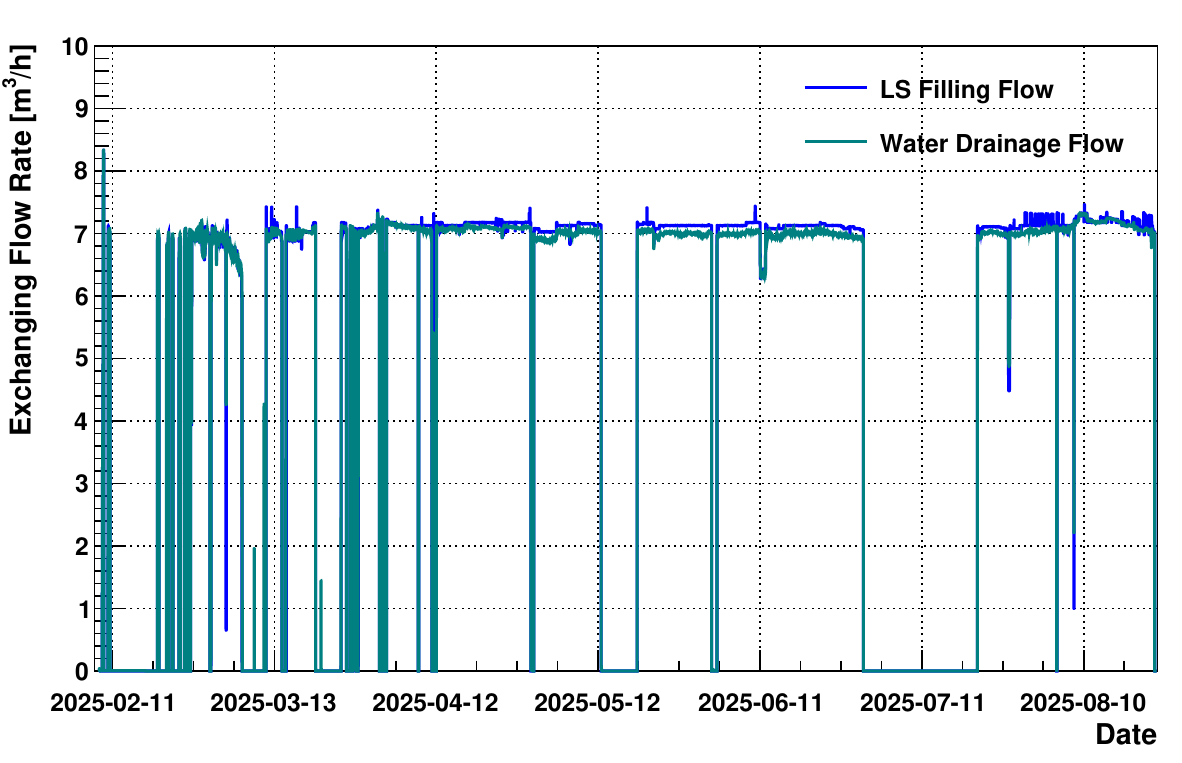}
    \captionof{figure}{Flow changes during the LS filling process}
    \label{fig:ex_flux}
\end{minipage}
\vspace{8pt} 

The LS filling process concluded with the liquid level in the CD stabilized at approximately 47 meters. At this liquid level, the detector exhibited stable and safe mechanical conditions, as validated by the continuous monitoring of the support rod forces throughout the entire exchange period. The measured forces on the rods showed general agreement with the FEA predictions within acceptable margins. The monitoring data confirmed that the maximum compressive force on the rods was around $-160\ \text{kN}$, and the maximum tensile force was around $99\ \text{kN}$. A few individual rods exhibited readings moderately higher than the FEA predictions, not exceeding 10\% of the maximum design values. The preliminary assessment attributes this deviation to potential factors such as the connection and installation details between the rods and the acrylic vessel. This aspect will continue to be monitored closely. Overall, the mechanical data indicate a very low risk of structural damage to the acrylic vessel or its support structure due to overstress under the current loading conditions.

Inevitably, during the extended stable operation period of the LS filling, several intermittent filling stops occurred. These were primarily attributed to two factors: temporary shortages of the raw LAB material from the supplier, and necessary adjustments and maintenance of the equipment. These intervals were utilized productively to conduct an intensive detector calibration campaign, which yielded valuable data for enhancing the understanding of the detector's response and refining subsequent data analysis parameters.

\subsection{LS Overflow and Circulation}
\label{subsec:ls-overflow}

\subsubsection{Control Strategy and Logic}
\label{sssec:overflow-control-logic}

During the long-term operation of the JUNO detector, the FOC system's role transitions from filling to managing the thermal dynamics of the large LS volume. The thermal volume expansion coefficient of LS ($8.8 \times 10^{-4}/^\circ\text{C}$) means that temperature fluctuations may cause significant volume changes—approximately \SI{20}{\cubic\meter} per \SI{1}{\celsius} change for the CD volume—which must be accommodated to maintain stable pressure conditions inside the sealed CD. The overflow system is specifically designed to handle these LS volume variations. The strategy for managing this employs a combination of passive and active mechanisms:

\begin{itemize}
    \item \textbf{Passive Thermal Regulation via Overflow Tanks}: The primary buffer for routine temperature fluctuations is provided by two dedicated overflow tanks. These tanks are connected to the CD's top chimney. For minor temperature variations, the resulting LS volume change is passively absorbed or replenished by these tanks through gravity-driven flow without any active intervention.
    
    \item \textbf{Active Volume Adjustment via Storage Tank}: For larger temperature shifts, the system switches to active mode. A separate storage tank is employed for this purpose. When the temperature drops significantly, valves and pumps are activated to transfer LS from the storage tank into the CD circuit to compensate for the contraction. Conversely, when the temperature rises significantly, the expanded LS volume automatically overflows into the overflow tanks. If needed, LS is actively transferred from the overflow tanks to the storage tank to prevent overfilling.

    \item \textbf{Online Purification Circulation}: To further reduce impurities over the detector's lifetime, the LS can be continuously purified through an online circulation loop. LS is extracted from the bottom of the CD, passed through external purification systems (e.g., water extraction, gas stripping), and then reinjected into the top of the CD. Computational Fluid Dynamics (CFD) simulations were employed to optimize this process~\cite{CD_paper}. The simulations revealed that the efficiency of purifying the entire CD volume depends on the temperature difference between the injected and in-situ LS. Without a temperature difference, one full CD volume exchange achieves only ~37\% circulation efficiency. By heating the purified LS to be $5^\circ\text{C}$ warmer than the CD LS, convection currents are suppressed, and the circulation efficiency increases to ~55\% per CD volume exchange. Based on these studies and the hydraulic characteristics, the optimal flow rate for this purification circulation was determined to be $7\ \text{m}^3/\text{h}$.
\end{itemize}

\subsubsection{Operational Implementation and Results}
\label{sssec:overflow-operation-results}

Following the completion of the LS filling on August 22nd, 2025, formal data acquisition for the JUNO experiment commenced on August 26th, 2025. The overflow system immediately became active to manage diurnal and other minor temperature variations.

The initial operation of the designated dedicated overflow pipeline encountered a technical challenge: an airlock phenomenon prevented the establishment of a reliable hydraulic communication between the overflow tanks and the top chimney of the CD. To promptly resolve this and ensure detector stability, the original LS filling pipeline was repurposed as a backup overflow pathway. This solution proved effective; the overflow process through this backup line operated smoothly, achieving overflow rates exceeding $2\ \text{m}^3/\text{h}$, which adequately met the initial operational requirements for handling thermal expansion.

However, to ensure mechanical safety, the final LS filling level was intentionally stabilized at approximately 47 meters, as validated by force monitoring data as mentioned in Section~\ref{sssec:ls-op}. This operational liquid level is below the original design value. Consequently, the liquid level in the interconnected FOC overflow tanks was also correspondingly lower. This reduction decreased the tanks' effective cross-sectional area for buffering level variations, reducing the system's passive overflow capacity to approximately 80\% of its nominal design value.

To maintain detector stability under this adapted configuration, two key measures were implemented:

The operational liquid level variation was constrained to within $\pm 5$ cm to minimize level-induced pressure variations on the acrylic vessel, a key requirement for mechanical safety. This liquid level tolerance is well within the high-precision control capabilities of the automated system, which can maintain level stability at this or even finer scales, thereby providing a robust safeguard for the detector's structural integrity.

An active liquid level adjustment function was integrated into the automatic control system to compensate for the reduced passive capacity. This system dynamically regulates the liquid level in the overflow tanks through small-volume transfers to and from the storage tank, maintaining precise balance with the CD's requirements. The control loop has been thoroughly validated and remains operational, ensuring long-term detector stability.

As observed in Figure~\ref{fig:overflow}, the liquid levels in the overflow tanks and the CD exhibit excellent synchronization. Currently, during the initial phase shortly after the completion of LS filling, the entire detector is undergoing a gradual cool-down. This results in a combined effect of thermal contraction in both the LS and the acrylic vessel itself, accompanied by elastic deformation of the spherical structure due to the changing liquid head. These factors collectively contribute to a progressive decrease in the liquid level. Under these conditions, the active replenishment function is activated, transferring LS into the CD to raise its liquid level slightly. This proactive compensation is crucial for maintaining the mechanical stability of the detector structure during this transitional phase.

\vspace{8pt} 
\noindent
\begin{minipage}{\columnwidth}
    \centering
    \includegraphics[width=1.0\columnwidth]{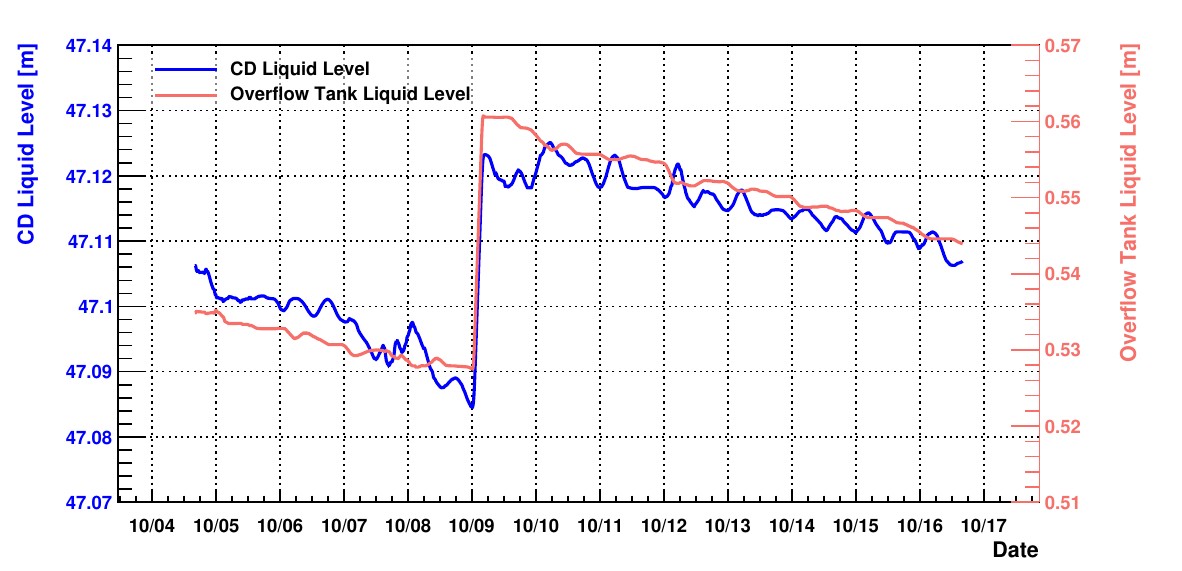}
    \captionof{figure}{Liquid Level Changes of Overflow Tanks and CD During JUNO Operation}
    \label{fig:overflow}
\end{minipage}
\vspace{8pt} 

\section{Summary}
\label{sec:conclusion}

The Filling, Overflow, and Circulation system has been successfully commissioned and has demonstrated its critical role in JUNO. Through meticulous design and robust automated control, the system safely and precisely managed the large-scale liquid handling operations, ensuring the structural integrity of the CD and preserving the exquisite optical and radiopurity properties of the LS. Its accomplishments can be summarized in three key aspects:

First, the FOC system has accomplished the filling of the world's largest liquid scintillator detector, taking 45 days for the initial water filling of approximately \SI{6.4e4}{tons} and 200 days for the subsequent LS filling of \SI{2.3e4}{\cubic\meter}, with 90\% operational availability during stable operation, while also ensuring the functionality of overflow management and future circulation.

Second, the integration of an automated control system based on PLC and EPICS software provided the foundation for this success, enabling real-time monitoring, precise control, and robust safety interlocking throughout all operations. The system has achieved fully automated liquid control, meeting stringent requirements for high-precision liquid level control within $\pm$2 cm and flow rate regulation within $\pm$0.5\% of setpoints.

Third, the design and operation of the FOC system have satisfied the demanding criteria of ultra-low background, minimal leakage, and long-term structural safety and reliability. The stringent radiopurity controls—including the use of ultra-high purity materials, meticulous cleanliness protocols, and exceptional leak tightness (helium leak rate \(<10^{-6}\) mbar-L\(\cdot\mathrm{s}^{-1}\) for assembled systems)—were instrumental in preserving the LS purity throughout the filling process. This foundational role is demonstrated by two key achievements: the maintenance of \(^{222}\)Rn concentration below 1 mBq/m\(^3\) in the fresh LS during filling, and the establishment of the conditions necessary for the outstanding ultimate radiopurity levels later confirmed by detector operation, which include \(^{238}\)U and \(^{232}\)Th concentrations below \(10^{-16}\) g/g and an initial \(^{210}\)Po rate of approximately \(5 \times 10^{4}\) cpd/kton. These collective results underpin the detector's potential for achieving its premier physics goals.

Looking forward, the established FOC infrastructure and operational expertise provide a solid foundation for future JUNO upgrades. Specifically, the system is designed to support potential next-phase physics programs, by enabling continuous online LS circulation to achieve even higher purity levels. Also it can replace LS for the research of neutrinoless double-beta decay.

In summary, the FOC system has proven to be a reliable, adaptable, and indispensable component for JUNO. Its successful implementation underscores the importance of integrated design, automated control, and continuous monitoring in large-scale neutrino detector operations, providing a solid foundation for JUNO's scientific exploration over its planned 20-year lifetime and beyond. The design principles, control strategies, and lessons learned from this project offer valuable insights for other large-scale, high-purity liquid-based detectors in particle physics and related fields.

\section*{Acknowledgement}

This work is supported by NO.2023YFA1606100, National Key R\&D Program of China.


\bibliographystyle{model1-num-names}

\bibliography{cas-refs}

\end{document}